\documentclass[letterpaper]{article} 
\usepackage{aaai25}  
\usepackage{times}  
\usepackage{helvet}  
\usepackage{courier}  
\usepackage[hyphens]{url}  
\usepackage{graphicx} 
\usepackage{natbib}  
\usepackage{caption} 
\usepackage{algorithm}
\usepackage{algorithmic}
\usepackage{multirow}
\usepackage{subcaption}
\usepackage{booktabs, multirow} 
\usepackage{soul}
\usepackage{subcaption,siunitx,booktabs}
\usepackage[english]{babel}
\usepackage{changepage,threeparttable}
\usepackage{placeins}

\usepackage{newfloat}
\usepackage{listings}
\usepackage{graphicx} 
\usepackage{subcaption}
\usepackage{footnote}
\usepackage{float}

\makesavenoteenv{tabular}
\DeclareCaptionStyle{ruled}{labelfont=normalfont,labelsep=colon,strut=off}
\floatstyle{ruled}
\newfloat{listing}{tb}{lst}{}
\floatname{listing}{Listing}

\title{Practical Guide for Causal Pathways and Sub-group Disparity Analysis}
\author{
    Farnaz Kohankhaki\textsuperscript{\rm 1}, Shaina Raza\textsuperscript{\rm 1}, Oluwanifemi Bamgbose\textsuperscript{\rm 1},\\Deval Pandya\textsuperscript{\rm 1}, Elham Dolatabadi\textsuperscript{\rm 1,2,*}, \\
}
\affiliations{
    \textsuperscript{\rm 1}Vector Institute,
    \textsuperscript{\rm 2}York University,\\
    
    \textsuperscript{\rm *}Corresponding Author: edolatab@yorku.ca
}

\usepackage{bibentry}

\begin{document}

\maketitle

\begin{abstract}
In this study, we introduce the application of causal disparity analysis to unveil intricate relationships and causal pathways between sensitive attributes and the targeted outcomes within real-world observational data. Our methodology involves employing causal decomposition analysis to quantify and examine the causal interplay between sensitive attributes and outcomes. We also emphasize the significance of integrating heterogeneity assessment in causal disparity analysis to gain deeper insights into the impact of sensitive attributes within specific sub-groups on outcomes. Our two-step investigation focuses on datasets where \textit{race} serves as the sensitive attribute. The results on two datasets indicate the benefit of leveraging causal analysis and heterogeneity assessment not only for quantifying biases in the data but also for disentangling their influences on outcomes. We demonstrate that the sub-groups identified by our approach to be affected the most by disparities are the ones with the largest ML classification errors. We also show that grouping the data only based on a sensitive attribute is not enough, and through these analyses, we can find sub-groups that are directly affected by disparities. We hope that our findings will encourage the adoption of such methodologies in future ethical AI practices and bias audits, fostering a more equitable and fair technological landscape.
\end{abstract}

\section{Introduction}
Fairness in data science and machine learning (ML) is indispensable for the \textit{responsible} development and deployment of ethical artificial intelligence (AI) technologies \citep{diaz2023connecting}. Key tools in data science, including Aequitas \citep{2018aequitas}, AI Fairness 360 \citep{aif360-oct-2018}, and Fairlearn \citep{bird2020fairlearn} play a pivotal role in addressing fairness challenges in ML models, focusing on concepts such as demographic parity and equalizing statistics across sensitive attribute groups \citep{saha2020measuring, feldman2015certifying, zhang2018equality, barocas-hardt-narayanan,raza2024fair}. However, these approaches can lead to fairness gerrymandering, where broad fairness across high-level groups masks unfair treatment within sub-groups \citep{kearns2018preventing}. Sub-group fairness approaches \citep{yang2020fairness, shui2022learning} have emerged to address this, aiming to reconcile group and individual fairness notions \citep{pfohl2023understanding}.

Furthermore, understanding and quantifying the extent to which the observed disparity in outcomes, such as those seen with demographic parity, is attributed to the causal influence of sensitive attributes is crucial in fields, including health and social sciences \citep{mehrabi2021survey,barocas2023fairness,braveman2011health,glymour2018causal}. Causality-based fairness frameworks view disparity as the causal effect of sensitive attributes $S$ on outcomes $Y$, raising fundamental questions about how changes in these attributes affect average outcomes \citep{kearns2018preventing}. These methodologies revolve around a central question: if the sensitive attribute $S$ changed (e.g., changing from marginalized group $s_1$ to non-marginalized group $s_2$), how would the outcome $Y$ change on average?

Two prominent causal frameworks, the structural causal model (SCMs) \citep{wu2019pc} and the potential outcome framework \citep{khademi2019fairness}, have been utilized for causal fairness analysis and more particularly to quantify the disparity \citep{mackinnon2007mediation,pearl2014interpretation}. SCMs assume that we have full knowledge of the causal graph, enabling us to decompose the causal effect of any variable into different paths, such as direct and indirect effects. On the other hand, the potential outcome framework \citep{rubin2005causal} does not assume the availability of the causal graph and instead focuses on estimating the causal effects of treatment variables. However, a common challenge across all causal models is identifiability, referring to whether they can be uniquely measured from observational data \citep{morgan2015counterfactuals}. This poses a critical barrier to applying these notions to real-world scenarios. 

Randomized experiments, considered the gold standard for inferring causal relationships in statistics, are often not feasible or cost-effective in the context of disparity analysis \citep{hariton2018randomised}. Therefore, in most cases, the causal relationship must be inferred from observational data rather than controlled experiments. This limitation has spurred a stream of research aiming to address these challenges and develop more practical and effective methodologies for causal fairness analysis. Early literature in the SCM primarily utilized linear and parametric methods, limiting its capacity to offer a universal approach for analyzing natural and social phenomena characterized by non-linearities and interactions \citep{mackinnon2007mediation}. Later, Pearl introduced the causal mediation formula designed for arbitrary non-parametric models, serving as a valuable tool for decomposing total effects \citep{pearl2014interpretation}. Subsequently, a substantial body of literature emerged, focusing on causal effect decomposition under the rubric of mediation analysis and proposing various optimization problems to adapt the causal framework for fairness analysis \citep{zhang2018fairness,wu2019pc,zhang2016causal}. One notable framework \citep{plecko2024causal} addresses spurious effects in the decomposition of causal effects and explores the relationships between causal and spurious effects with demographic parity, offering practical insights for data science and fairness considerations. 

In the realm of fairness through causal analysis research, a significant focus lies on sub-group analysis and heterogeneity, approached from two perspectives: one being heterogeneous treatment effects \citep{wager2018estimation}, which directly aligns with our study, and the other involving 'counterfactually fair' algorithms for individuals, a topic not directly relevant to our current research \citep{kearns2019empirical, kavouras2024fairness}. The former involves systematically quantifying variations in the causal impact of sensitive attributes on the outcome of interest across sub-groups \citep{pearl2022detecting}. Approaches for estimating heterogeneous causal effects encompass classical non-parametric methods such as nearest-neighbour matching, kernel methods, and series estimation, demonstrating efficacy in scenarios with a limited number of covariates \citep{crump2008nonparametric, lee2009non, willke2012concepts}. More recently, data-driven ML algorithms including causal forest which can be adept at handling numerous moderating variables have shown promising results in heterogeneity analysis \citep{wager2018estimation}. 

Building on the urgency of adopting causal reasoning techniques in fairness analysis, the main aim of this study is to leverage causal analysis for sub-group disparity assessment. First, we demonstrate the application of causal disparity analysis to uncover the intricate relationships and causal pathways between sensitive attributes and the outcome of interest in real-world observational data. Then, we close the loop by employing causal disparity analysis for sub-group fairness within the context of ML, showcasing how a causal-aware approach can enhance sub-group fairness evaluation. Our overarching goal is to pave the way for conducting disparity audits that lay the foundation for ethical and equitable ML. The novelty of this study lies not in the specific methodologies used but in recognizing causal reasoning as a novel technique for conceptualizing and quantifying disparity, making it suitable for promoting fairness in data science.
\begin{itemize}
    \item We demonstrate the application of causal disparity analysis to quantify and decompose causal pathways between sensitive attributes and the targeted outcomes within two real-world observational data. We successfully indicate the capability of our approach to uncover hidden disparities, even in cases where observed disparities are nearly zero. 
    \item We pioneer a novel sub-group discovery method rooted in the concept of \textit{Heterogeneity of Treatment Effect}, enabling the identification of variations in the magnitude and direction of decomposed causal effects among individuals.
    \item We evaluate the efficacy and utility of our proposed causal disparity analysis in a fairness ML experiment. Our method demonstrates its ability to identify biased performance within each sub-group of individuals, particularly those identified quantitatively as most affected by disparities.
\end{itemize}

\section{Materials and Methods}
In this section, we will introduce causal disparity analysis through the lens of counterfactual inference and non-parametric SCM proposed by Pearl \citep{pearl2014interpretation} and expanded by Zhang et al. \citep{zhang2018fairness}. Following this approach, various causal effects can be defined as the difference between two counterfactual outcomes \citep{holland, rubin1974estimating} along the causal pathway from sensitive attributes (causes) to outcomes. We will elucidate how these effects can be quantitatively measured and estimated from data through the experiments. 

\subsection{Preliminaries}
Our study is based on a basic causal structure which consists of four random variables $(Y, S, X, M)$ sampled from unknown distribution; $S$ represents the random variable for the sensitive attribute (whose effect we seek to measure). $Y$ represents the random variable for the outcome of interest. $X$ represents the random variable for all in-sensitive attributes, including observed confounders, denoted by $C$, and mediators, denoted by $M$. The lowercase $(y, s, x, m)$ represents the values that variables may take. As a running example, $S$ stands for \textit{race}, $M$ stands for the job title, and $Y$ stands for income amount. Here we consider two potential outcomes $Y_{s1}$ and $Y_{s2}$ for sensitive attributes, $S={s_1,s_2}$. $E[Y_s,m]$ stands for $E[Y|do(S=s, M=m)]$ which is interpreted as the expectation of potential outcome $Y$ when the sensitive attribute $S$ is set to $s$ and the mediator variable $M$ is set to $m$. 

Sensitive attributes, $S$, that serve as the basis for disparity encompass a range of personal characteristics that have historically been unfairly targeted to differentiate individuals \citep{mehrabi2021survey}. These attributes are pivotal in discussions surrounding equity, inclusion, and human rights and are commonly discussed in anti-discrimination laws \cite{human_rights}, regulations, and human rights frameworks around the globe. Among these attributes, a notable array includes \textit{race}, nationality, ethnic origin, colour, religion, age, sex, sexual orientation, gender identity or expression, marital status, family status, genetic characteristics, or disability \citep{verma2018fairness}.

In this study, the term \textit{sensitive category} denotes individuals grouped based on their sensitive attributes. The term \textit{sub-group} refers to individuals grouped according to the quantity of their estimated causal effects.

\subsection{Causal Disparity Analysis} 
Within the context of counterfactual fairness, the causal effect is characterized as the difference between two potential (also called counterfactual) outcomes: one outcome, $Y_{s_1}$, if the sensitive attribute is $s1$ (for instance, if the individual is female), and the other outcome, $Y_{s_2}$ for $s2$ (in this case, if the individual is not female). Due to the presence of the mediator, the potential outcomes are not only dependent on sensitive attributes but also on mediator values [4]. This way the causal effect can be decomposed into effects such as counterfactual causal effect, counterfactual indirect effect and spurious effect. The counterfactual measures of direct and indirect effects, are conditional versions of the natural direct and indirect effect introduced by Pearl~\cite{pearl2014interpretation} and are widely popular throughout the empirical sciences. Here we define the causal and non-causal fairness criteria we have used in this study;

\textbf{Total Variation (TV)} also known as demographic parity represents the statistical distinction in the conditional distribution of the outcome between two groups when simply observing that $S = s_1$, compared to $S = s_2$:
\begin{equation}
    TV(Y) = P(Y|S = s_2)- P(Y|S = s_1)
\end{equation}

The \textbf{counterfactual direct effect (ctf-DE)} is the average difference between two potential outcomes when the sensitive attribute transitions from $S = s_1$ (female) to $S = s_2$ (not female), while the mediator is set to whatever value it would have naturally attained prior to the change in $S = s_1$, for a specific sub-group of the population, $s$.  
\begin{equation}
    ctf-DE_{s_1,s_2} (Y|s)= E[Y_{s_2},M_{s_1}- Y_{s_1},M_{s_1}|s]
\end{equation}
 
The \textbf{counterfactual indirect Effect (ctf-IE)} is the average difference between two potential outcomes when the sensitive attribute remains constant at $s_1$ (female), while the mediator changes from its values under $s_1$ to whatever value it would have attained for each individual under $s_2$ (not female), for a specific sub-group of the population, $s$.

\begin{equation}
ctf-IE_{s_1,s_2}(Y|s)= E[Y_{s_1},M_{s_2}- Y_{s_1},M_{s_1}|s]
\end{equation}

According to Zhang~\cite{zhang2018fairness}, direct and indirect causal effects can be linearly combined and contribute to total variation by introducing an additional term that uncovers the spurious relations between $S$ and $Y$ through confounding variables, $X$. 
\begin{equation}
    TV_{s_1,s_2}(Y) = DE_{s_1,s_2} (Y|s) - IE_{s_2,s_1}(Y|s) - SE_{s_2,s_1}(Y) 
\end{equation}

\textbf{Counterfactual Spurious Effect (ctf-SE)} measures the average difference in outcome $Y$ had $S$ been $s_1$ by intervention compared to settings that would naturally choose $S$ to be $s_1$. SE, in fact, measures all paths between $S$ and $Y$ except the causal ones (direct and indirect),
\begin{equation}
    ctf-SE_{s_1,s_2}(Y|s)= E[Y_{s_1}|s_2- Y_s|s_1]
\end{equation}
In order to estimate these counterfactual quantities from data, we assume the presence of unconfoundedness between the sensitive attribute and outcome, along with the assumption of conditional ignorability. Moreover, leveraging the following two assumptions (1) none of the confounders are descendants of $S$ and (2) confounders block all backdoor paths from mediators to $Y$, we can express counterfactual quantities in terms of conditional distributions. 

\subsection{Sub-group discovery for heterogeneity assessment} 
We conduct sub-group discovery to identify and quantify causal effect heterogeneity among individuals from distinct sensitive categories (sensitive attributes are changed while keeping all other relevant variables constant). Generalized Random Forest (GRF) \cite{athey2019estimating} is employed in this study to measure conditional distribution which is an extension of traditional Random Forest by maximizing heterogeneity when splitting nodes in a decision tree. It incorporates a statistical criterion known as the Causal tree-splitting criterion, which integrates sensitive attribute assignments and outcome variables. GRF provides estimates of both average and individual causal effects, facilitating the detection of differential effects among sub-groups. This allows for the clustering and grouping of individual effects to reveal varying causal impacts. Essentially, GRF compares individuals within a sub-group to counterparts with different sensitive attributes while aiming to closely match all other relevant attributes.

\subsection{Experiment and Setting}
\begin{figure}[ht]
  \centering
  \includegraphics[width=0.8\linewidth]{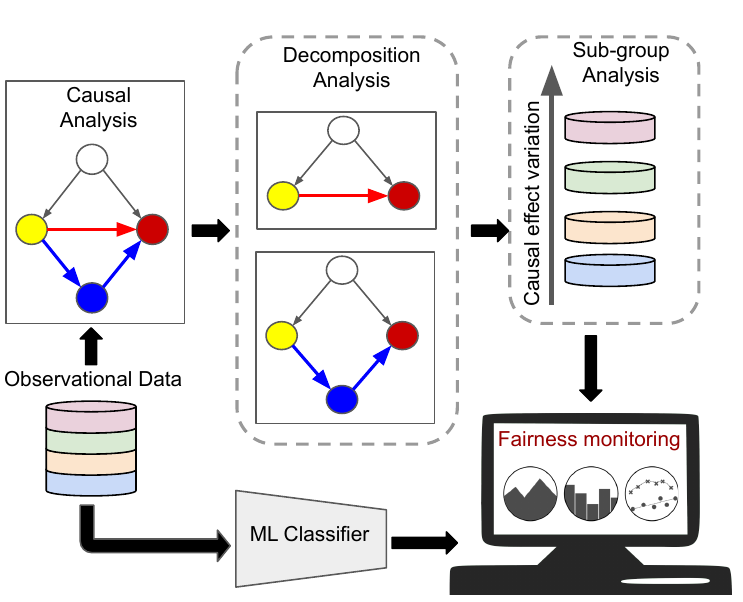}
  \caption{The steps involved in our approach to achieving fairness in ML classification models through causal pathway decomposition and sub-group analysis.}
  \label{fig:pipeline}
\end{figure}

We have leveraged causal and sub-group analysis for disparity analysis using the pipeline shown in Figure \ref{fig:pipeline} on two publicly available datasets, where we selected \textit{race} as the sensitive attribute. We identified individuals with one \textit{race} as the $s_1$ group and the rest as the $s_2$ group. Please refer to Table~\ref{tab:dataset} for more details on the attributes designated as confounders, mediators, and outcomes. Within our pipeline, we utilized the \textit{faircause} library \cite{plecko2022causal} and \textit{GRF} \cite{wager2018estimation} for causal effect estimations.

\begin{table}   [ht]
\centering
\caption{The description of variables and their corresponding nodes in our causal graph for the disparity analysis experiments. where the sensitive attribute is defined as the White vs. non-White (Asian vs. Non-Asian) \textit{race}.}
\label{tab:dataset}
\resizebox{0.5\textwidth}{!}{
\begin{tabular}[]{lll}
\hline
\multicolumn{1}{l}{\multirow[t]{2}{*}{\textbf{Dataset}}} & \multicolumn{2}{c}{\textbf{Attributes}}  \\
 & \textbf{Node} & \textbf{Types and Variables} \\ \hline
Adults & Sensitive (S)                     & Race (S1: Non-White, S2: White) \\
       & Outcome (Y)                       & Salary (2 categories: $\le$\$50K/\textgreater{}\$50K) \\
       & Confounders (X)                   & \begin{tabular}[t]{@{}l@{}}Age (continuous),  \\ Applicant Sex (2 categories), \\ Marital Status (7 categories)\end{tabular}\\
                                                      & Mediators (M)                     & \begin{tabular}[t]{@{}l@{}}Education (16 categories)\\ Workclass (8 categories)\\ Occupation (14 categories)\\ Capital Gain (continuous)\\ Capital Loss (continuous)\\ Hours/Week (continuous)\end{tabular} \\ \hline
\multirow[t]{5}{*}{HDMA}                                 & Sensitive  (HDMA-White)                       & Race (S1: Non-White, S2: White)\\
   & Sensitive (HDMA-Asian)                        & Race (S1: Asian, S2: Non-Asian)\\
                                                      & Outcome                           & Loan Status (Accepted/Rejected) \\
                                                      & Confounders                       & \begin{tabular}[t]{@{}l@{}}Property Type (2 categories)\\ Owner Occupancy (3 categories) \\ Applicant Sex (2 categories) \\ Loan Type (4 categories)\end{tabular}                                                       \\
                                                      & Mediators                         &    \begin{tabular}[t]{@{}l@{}}Loan Amount (continuous)\\ Applicant Income (continuous) \end{tabular}  \\ \hline
\end{tabular}}
\end{table}

\textbf{Adult}. The adult dataset \cite{adult} is a multivariate dataset designed to predict whether an individual's annual income will exceed \$50,000. This prediction is based on census data and is commonly known as the 'Census Income' dataset. The data extraction was carried out by Barry Becker, utilizing the 1994 Census database. In this dataset, the goal is to identify and quantify the basic impact of individuals’ \textit{race} (specifically white) on income as listed in detail in Table \ref{tab:dataset}.

\textbf{HDMA}. The Home Mortgage Disclosure Act (HMDA) \cite{hdma} mandates numerous financial institutions to uphold, report, and openly divulge mortgage-related information. These publicly accessible data hold significance as they provide insights into whether lenders are effectively addressing their communities' housing requirements. They also furnish public officials with valuable information to facilitate decision-making and policy formation, while also unveiling lending trends that could potentially exhibit bias. For our experiments, we leveraged the HDMA “Washington State Home Loans, 2016" dataset comprising a total of 466,566 instances of home loans within the state of Washington. The variables encompass a diverse range of information, including demographic details, location-specific data, loan status, property and loan types, loan objectives, and the originating agency. For HDMA, we conducted two sets of experiments, referred to as HDMA-White and HDMA-Asian to explore the impact of the presence and absence of specific \textit{race}s on the outcome. Please refer to Table~\ref{tab:dataset} for more details.

\section{Results}
\subsection{Causal aware disparity analysis}
In Table~\ref{tabcfa}, we present total variations along with decomposed causal effects using causal forest for experimental datasets. All metrics are computed based on the difference in outcomes when the sensitive attribute transitions from $s_1$ to $s_2$. Positive results for our experiment favour individuals with $s_2$, while negative results favour the other sensitive group, which is $s_1$. Furthermore, we conducted comparisons of the causal effect estimates in our pipeline with two other widely-used causal decomposition libraries, as illustrated in Supplementary Table \ref{tab:natural-effects}. 

\begin{table}[ht]
\caption{Total Variation (TV), Direct Effect (DE), Indirect Effect (IE), and Spurious Effect (SE) estimations for each experiment.}
\label{tabcfa}

\begin{tabular}{lllll}
\hline
\hline
\textbf{Dataset}                                                       & \textbf{TV}\footnote{Total Variation} & \textbf{ctf-DE}\footnote{Direct Effect} & \textbf{ctf-IE}\footnote{Indirect Effect} & \textbf{ctf-SE}\footnote{Spurious Effect} \\
\hline
\textbf{Adult}                                                         & 0.104       & 0.015       & 0.032       & 0.057        \\
\hline  
\textbf{\begin{tabular}[t]{@{}l@{}}HDMA-White\end{tabular}} & 0.041       & 0.055       & -0.005      & -0.009       \\
\hline  
\textbf{\begin{tabular}[t]{@{}l@{}}HDMA-Asian \end{tabular}} & 0.005      &     0.030	& -0.009	& -0.017               \\
\hline 
\end{tabular}

\end{table}

Our findings reveal that within the \textit{Adult} dataset, individuals from $s_1$ group are approximately 10.4\% more inclined to obtain an annual income exceeding \$50,000 than the $s_2$ group. Through causal analysis, we discern that about 1.5\% of this 10.4\% can be directly attributed to the causal influence of the sensitive attribute on the annual income (the full distribution of the ctf-DE is shown in the supplementary materials Figure \ref{fig:ctf_de_hist}). Additionally, approximately 3.2\% can be attributed to an indirect effect mediated through other factors shown in table~\ref{tab:dataset}, while the remaining 6\% is attributable to spurious effects.

In both experiments conducted with the \textit{HDMA} dataset, minimal disparities were observed through TV. Specifically, there was a mere 4\% difference and nearly zero TV between the two sensitive groups in loan acceptance status. In the first HDMA experiment, featuring a 4\% TV, a 5.5\% direct effect from the \textit{race} to loan status was noted, while both the indirect and spurious effects were negligible. However, the analysis of the second experiment (last row in Table~\ref{tabcfa}) yielded intriguing results. Despite the absence of TV and an indirect effect, there was a 3\% direct effect observed alongside an approximately 1.7\% negative spurious effect from the relevant factor to loan status. Additionally, as shown in the supplementary materials (Figures 5 and 6), the full distribution of direct causal effects for both HDMA experiments is skewed positively in favour of White and non-Asian sub-populations.

\begin{figure*}
\begin{subfigure}{.33\textwidth}
    \centering
    \includegraphics[width=.99\linewidth]{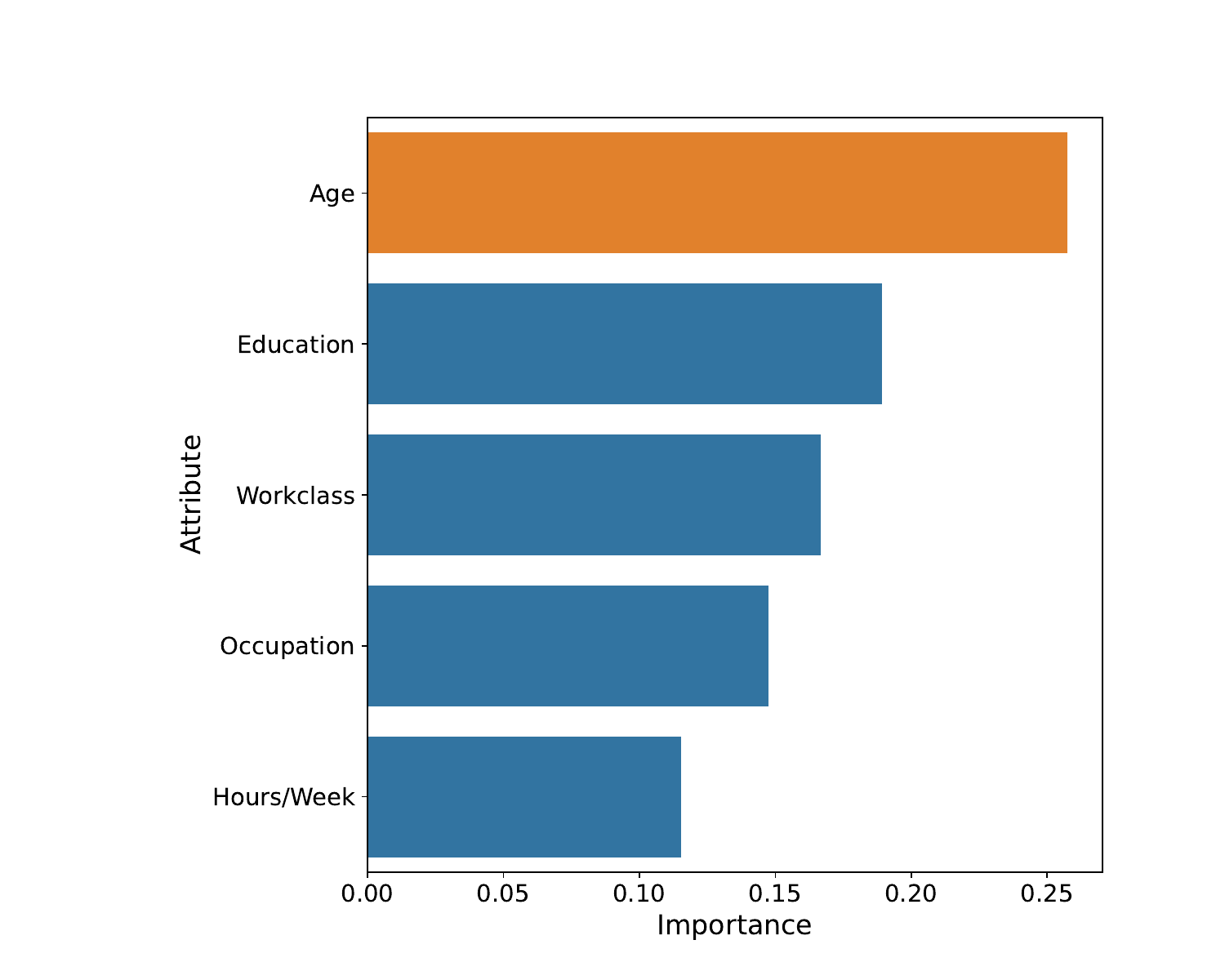}  
    \caption{Adult}
    \label{fig:var_imp_adult}
\end{subfigure}
\begin{subfigure}{.33\textwidth}
    \centering
    \includegraphics[width=.99\linewidth]{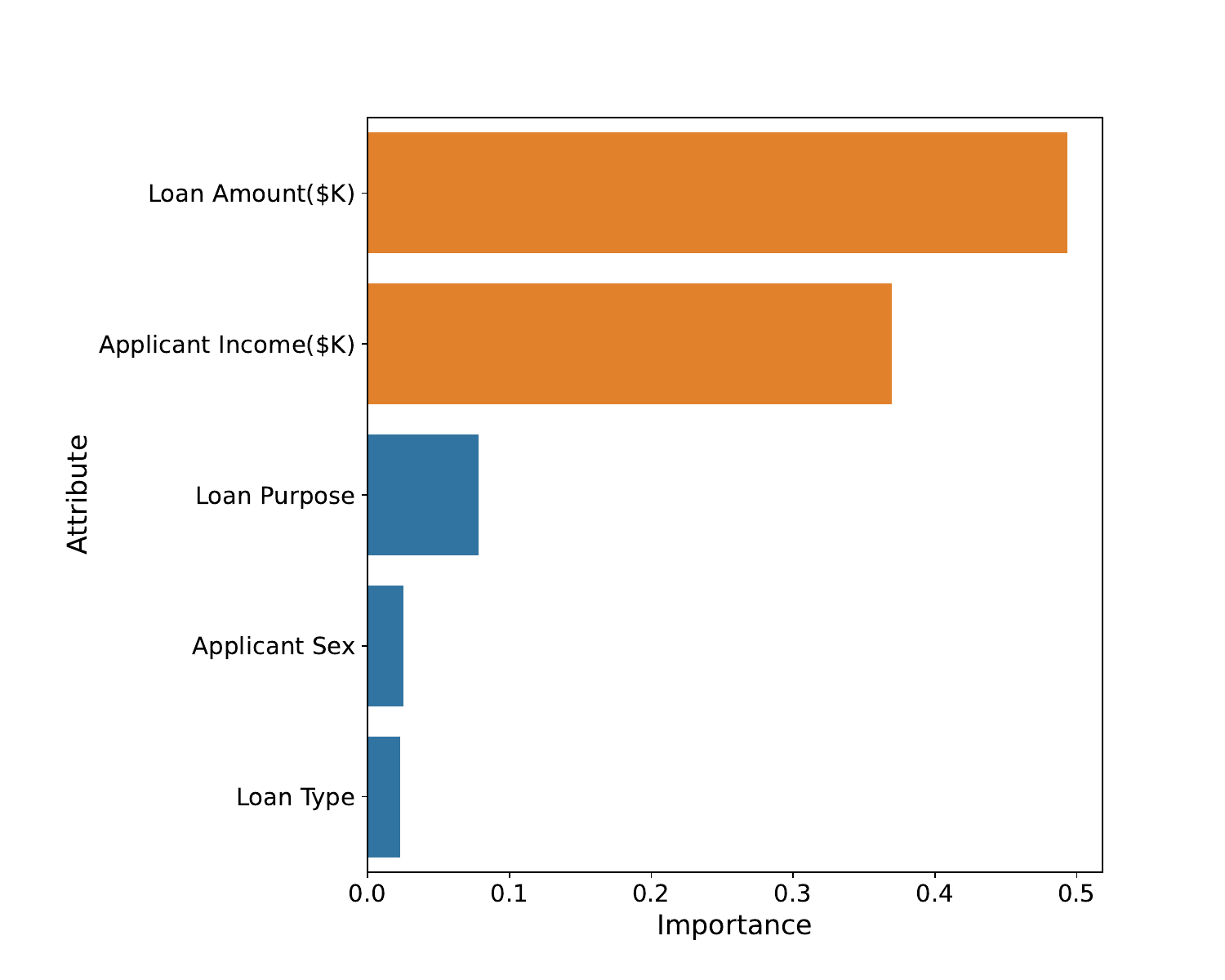} 
    \caption{HDMA-White}
\end{subfigure}
\begin{subfigure}{.33\textwidth}
    \centering
    \includegraphics[width=.99\linewidth]{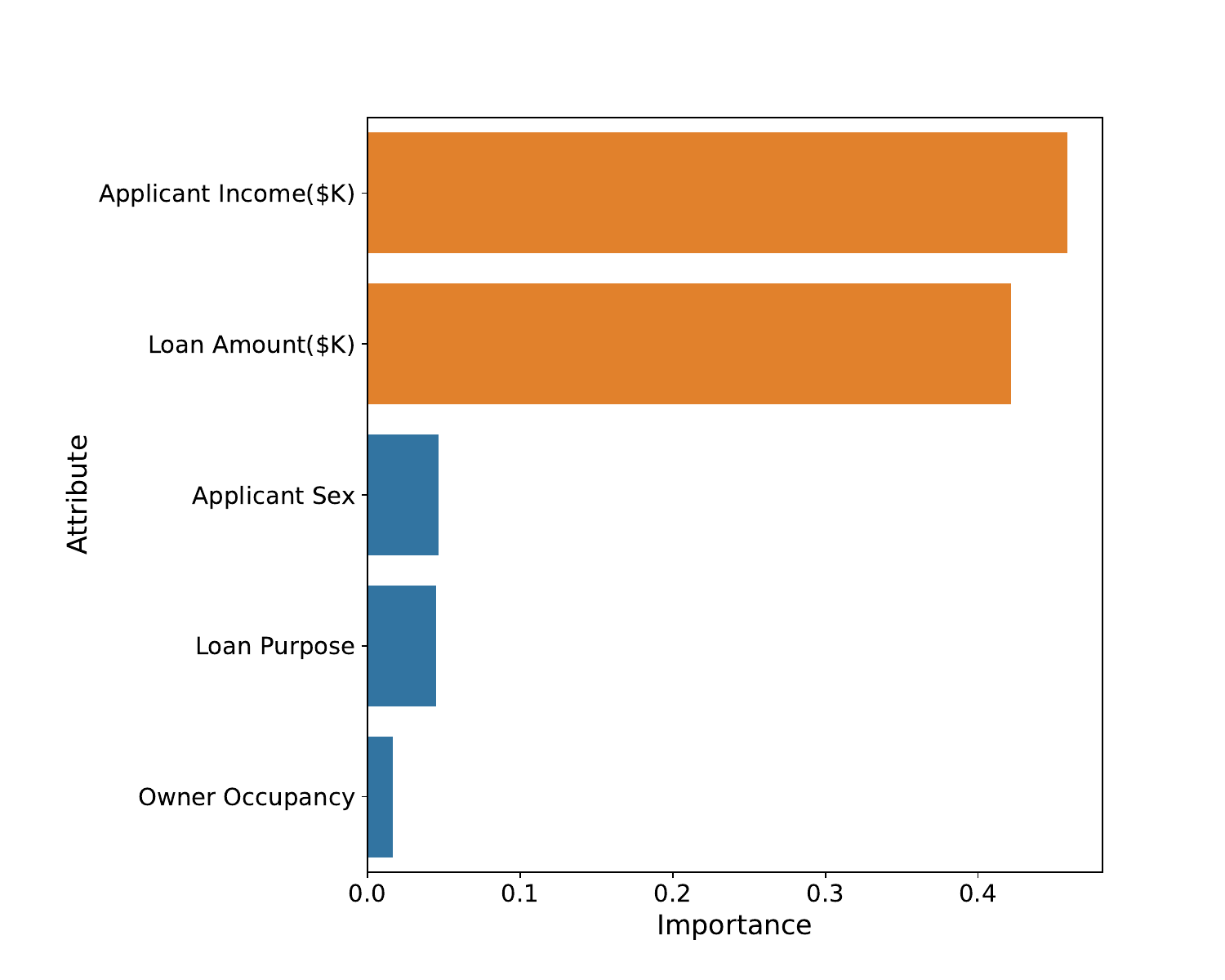}  
    \caption{HDMA-Asian}
\end{subfigure}
\caption{Variable importance for top 5 attributes of each experiment.}
\label{fig:var_imp}
\end{figure*}

Figure~\ref{fig:var_imp} presents the top attributes identified within each experimental setting for direct causal effect estimation. In the Adult dataset, key attributes include age, education, workclass, occupation, and hours per week,while for HDMA, loan amount and application income are pivotal.

\subsection{Sub-group analysis}
\begin{figure*}
\centering
\begin{subfigure}{.4\textwidth}
    \centering
    \includegraphics[width=.95\linewidth]{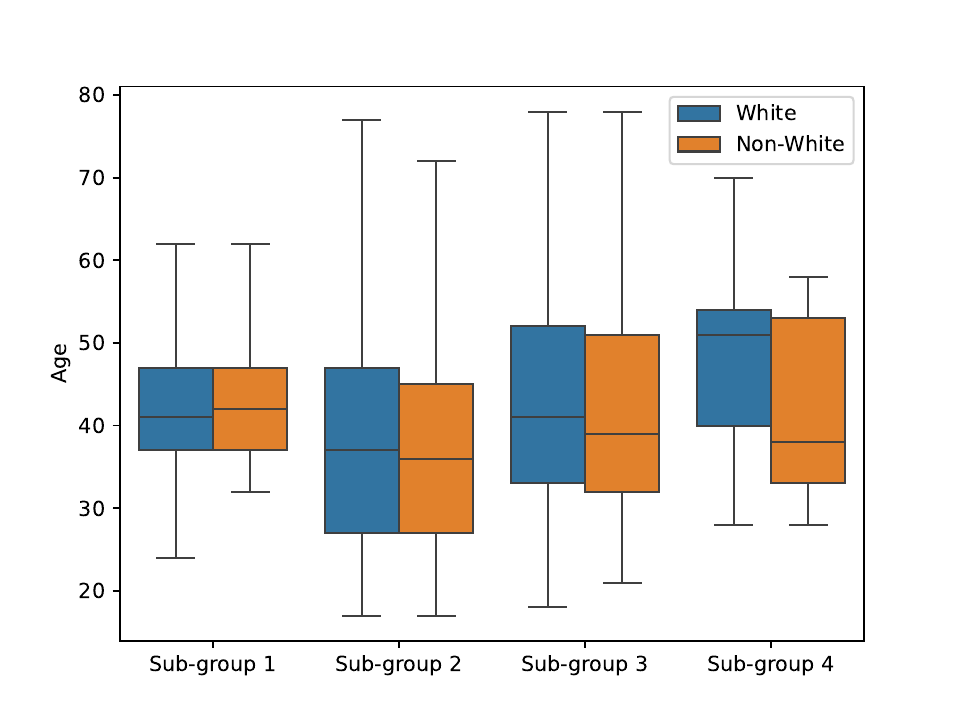}  
     \caption{dd}
     
\end{subfigure}
\begin{subfigure}{.4\textwidth}
    \centering
    \includegraphics[width=.95\linewidth]{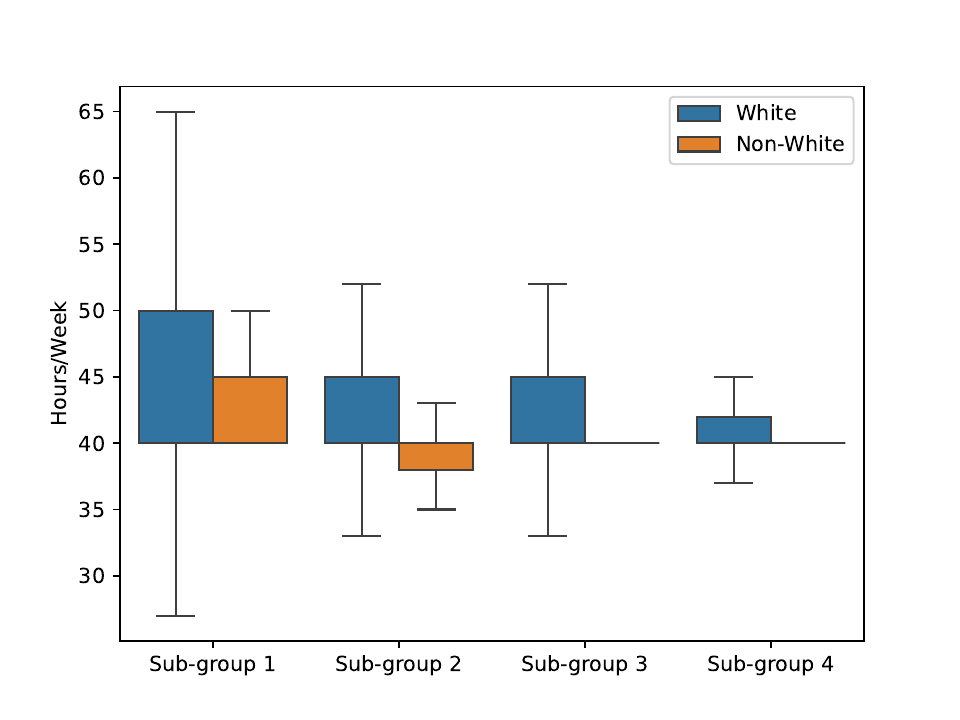}  
     \caption{}
     
\end{subfigure}

\begin{subfigure}{.4\textwidth}
    \centering
    \includegraphics[width=.95\linewidth]{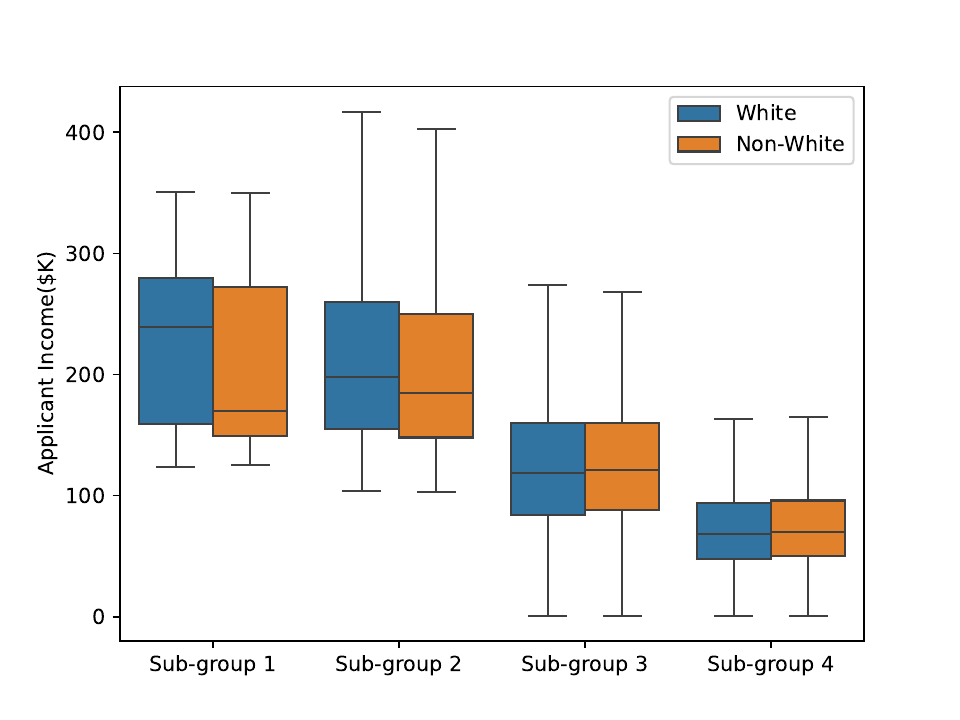}  
     \caption{}
     
\end{subfigure}
\begin{subfigure}{.4\textwidth}
    \centering
    \includegraphics[width=.95\linewidth]{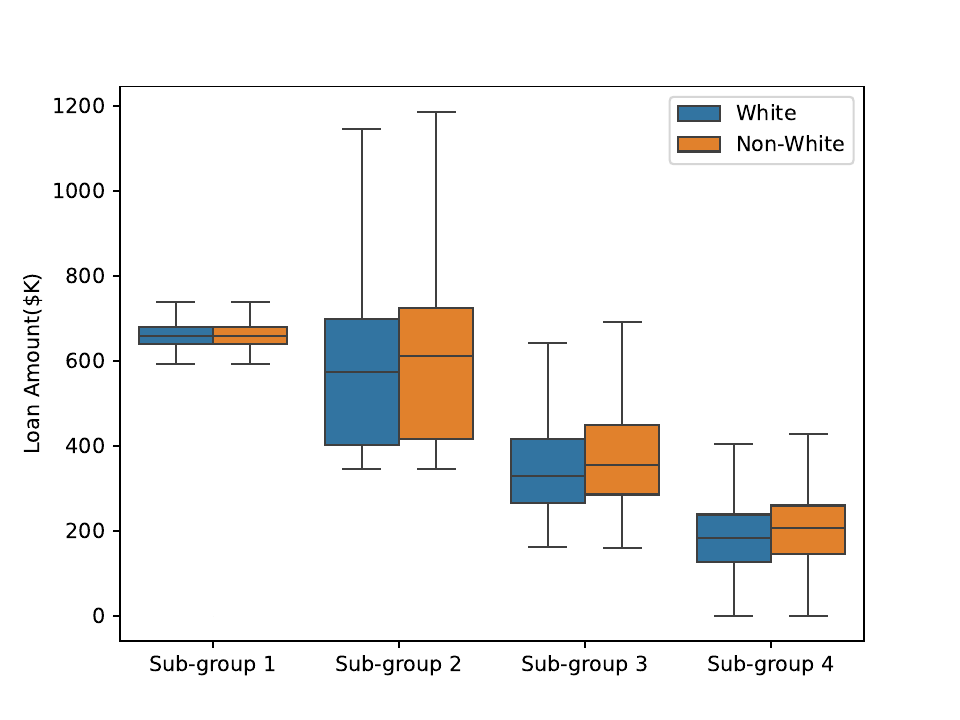}  
     \caption{}
     
\end{subfigure}

\begin{subfigure}{.4\textwidth}
    \centering
    \includegraphics[width=.95\linewidth]{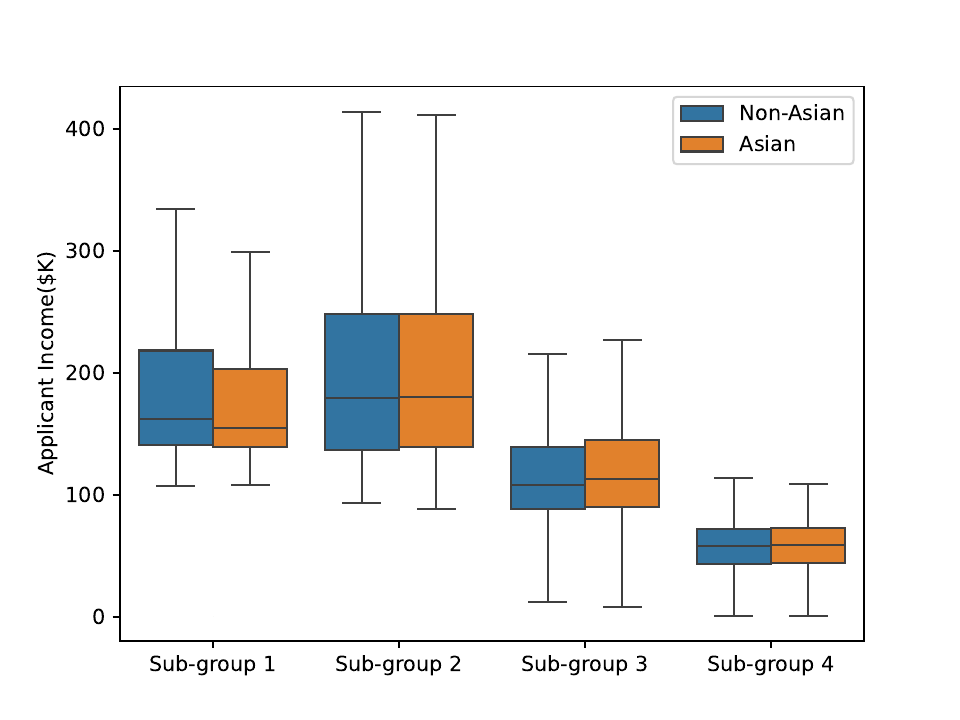}  
     \caption{E}
     
\end{subfigure}
\begin{subfigure}{.4\textwidth}
    \centering
    \includegraphics[width=.95\linewidth]{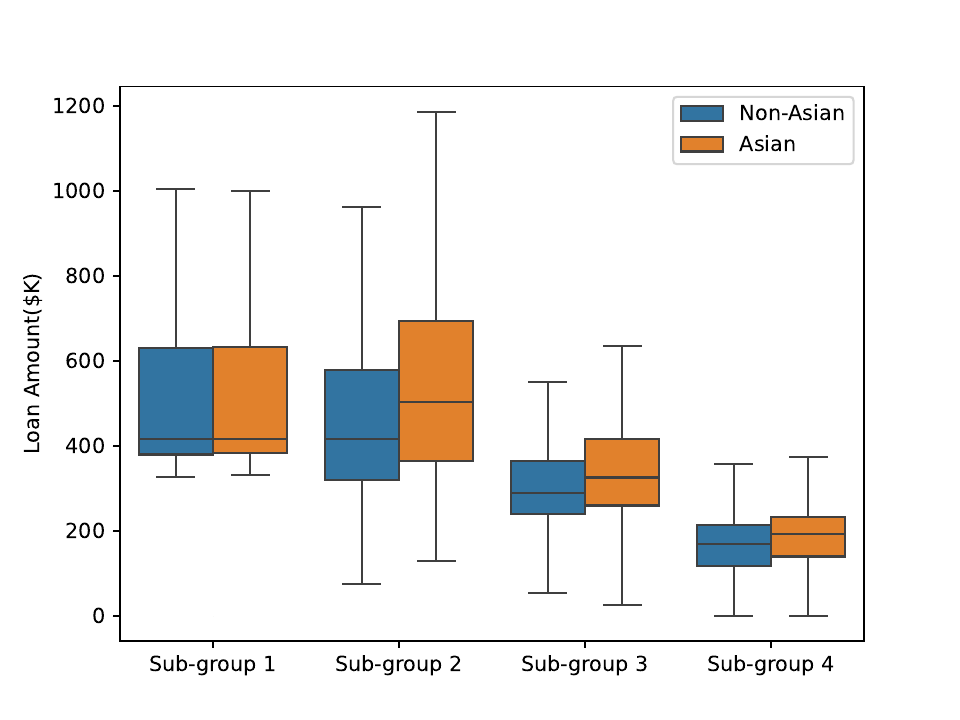}  
    \caption{}
     
\end{subfigure}

\caption{Summary of sub-group analysis for the continuous Variables: (a) and (b) represent \textit{Age} and \textit{Hours per Week} in the Adult dataset; (c) and (d) represent \textit{Application Income} and \textit{Loan Amount} in the HDMA-White dataset; and (e) and (f) represent \textit{Application Income} and \textit{Loan Amount} in the HDMA-Asian dataset. Sub-group 1 represents \textit{ctf-DE} values less than $-0.01$, Sub-group 2 represents \textit{ctf-DE} values between $-0.01$ and $0.01$ (around zero effects), Sub-group 3 represents \textit{ctf-DE} values between $0.01$ and $0.05$, and Sub-group 4 represents \textit{ctf-DE} values greater than $0.05$.}
\label{fig:adult-vars}
\end{figure*}

Drawing on the distribution of the ctf-DE, as shown in the supplementary materials (Figure \ref{fig:ctf_de_hist}), we examined the trade-off between ensuring consistency across datasets with varying distributions and maintaining intra-group alignment on ctf-DE ranges for each dataset to minimize variations. We, therefore, determined four distinct sub-groups with the summary shown in Table~\ref{tab-adult} for categorical variables and Figure~\ref{fig:adult-vars} for continuous variables, across all two experimental datasets. The sub-groups are arranged from negative to positive direct causal effect values, with Sub-group 1 representing ctf-DE values less than $-0.01$ (negative effects in favour of individuals in $s_1$ category), Sub-group 2 comprising ctf-DE values between $-0.01$ and $0.01$ (around zero effects), Sub-group 3 encompassing ctf-DE values between $0.01$ and $0.05$ (positive effects in favour of individuals in $s_2$ category), and Sub-group 4 indicating values greater than $0.05$ (very positive effects in favour of individuals in $s_2$ category) for the \textit{Adult} dataset. For the \textit{HDMA} dataset, the sub-groups are slightly different due to the ctf-DE values being skewed positively. Sub-group 1 has ctf-DE values less than $-0.005$, Sub-group 2 ctf-DE values between $-0.005$ and $0.025$, Sub-group 3 ctf-DE values between $0.025$ and $0.07$ (in favour of White or non-Asian), and Sub-group 4 indicating values greater than $0.07$. For the categorical variables, we have reported the counts for the majority and non-majority categories for each sensitive (racial) group within each of the sub-groups. Evidently, there are remarkable similarities in both majority and minority counts and mean and standard deviation between the two sensitive categories within each sub-group. 

\begin{table*} [ht]
\caption{Summary of sub-group analysis for the Adult experiment with categorical variables. Sub-group 1 represents \textit{ctf-DE} values less than $-0.01$, Sub-group 2 represents \textit{ctf-DE} values between $-0.01$ and $0.01$ (around zero effects), Sub-group 3 represents \textit{ctf-DE} values between $0.01$ and $0.05$, and Sub-group 4 represents \textit{ctf-DE} values greater than $0.05$.}

\centering
\resizebox{\textwidth}{!}{
\begin{tabular}{llllllll}
\hline
\toprule
 &  & \multicolumn{2}{c}{\textbf{Education-Num}} & \multicolumn{2}{c}{\textbf{Workclass}} & \multicolumn{2}{c}{\textbf{Occupation}} \\ \cmidrule{3-8}
 &  & \textbf{Majority} & \textbf{Minority} & \textbf{Majority} & \textbf{Minority} & \textbf{Majority} & \textbf{Minority} \\ \midrule
 \multicolumn{8}{l} {\textbf{Adults}}\\ \midrule
\textbf{Sub-group 1} & White & College (\%57.6) & Preschool (\%0.2) & Private (\%86.1) & State Gov. (\%0.5) & Craft Repair (\%43.1) & Transportation (\%0.4) \\
\textbf{\textbf{(TV:0.04)}} & Non-White & College (\%57.7) & Masters (\%3.8) & Private (\%88.5) & Self-emp.(NI) (\%3.8) & Craft Repair (\%53.8) & MOI \footnote{Machine Operator Inspection} (\%1.9) \\ \midrule
\textbf{Sub-group 2} & White & High School (\%32.2) & Preschool (\%0.2) & Private (\%73.9) & Without Pay (\%0.1) & Craft Repair (\%13.0) & Armed-Forces (\%0.0) \\
\textbf{\textbf{(TV:0.09)}} & Non-White & High School (\%34.4) & Preschool (\%0.3) & Private (\%74.5) & Without Pay (\%0.0) & Other Service (\%18.2) & Armed-Forces (\%0.0) \\ \midrule
\textbf{Sub-group 3} & White & High School (\%38.3) & 1st-4th Grade (\%0.1) & Private (\%69.8) & Without Pay (\%0.0) & Prof. Specialty (\%20.0) & Armed-Forces (\%0.0) \\
\textbf{\textbf{(TV:0.12)}} & Non-White & High School (\%37.0) & 11th Grade (\%0.2) & Private (\%68.5) & Self-emp.(I) (\%3.0) & Prof. Specialty (\%23.2) & Armed-Forces (\%0.2) \\ \midrule
\textbf{Sub-group 4} & White & Bachelors (\%50.7) & Assoc. Voc (\%0.2) & Private (\%70.0) & Local Gov. (\%0.2) & Exec-managerial (\%40.1) & Protective Serv. (\%0.4) \\
\textbf{\textbf{(TV:0.1)}} & Non-White & Bachelors (\%37.7) & College (\%3.8) & Private (\%66.0) & Federal Gov. (\%1.9) & Prof. Specialty (\%49.1) & MOI (\%1.9)\\ \midrule

 &  & \multicolumn{2}{c}{\textbf{Loan Purpose}} & \multicolumn{2}{c}{\textbf{Applicant Sex}} & \multicolumn{2}{c}{\textbf{Loan Type}} \\ \cmidrule{3-8}
 &  & \textbf{Majority} & \textbf{Minority} & \textbf{Majority} & \textbf{Minority} & \textbf{Majority} & \textbf{Minority} \\ \midrule
  \multicolumn{8}{l} {\textbf{HDMA - White}}\\ \midrule
\textbf{Sub-group 1} & White & Refinancing (\%77.0) & Home Improv. (\%1.3) & Male (\%100.0) & Male (\%100.0) & Conventional (\%100.0) & Conventional (\%100.0) \\
\textbf{\textbf{(TV:0.05)}} & Non-White & Refinancing (\%73.7) & Home Purchase (\%26.3) & Male (\%100.0) & Male (\%100.0) & Conventional (\%100.0) & Conventional (\%100.0) \\ \midrule
\textbf{Sub-group 2} & White & Refinancing (\%63.9) & Home Improv. (\%2.2) & Male (\%93.6) & Female (\%6.4) & Conventional (\%98.8) & FHA-insured (\%0.4) \\
\textbf{\textbf{(TV:-0.02)}} & Non-White & Refinancing (\%49.6) & Home Improv. (\%1.0) & Male (\%93.7) & Female (\%6.3) & Conventional (\%99.6) & FHA-insured (\%0.1) \\ \midrule
\textbf{Sub-group 3} & White & Home Purchase (\%69.2) & Home Improv. (\%2.6) & Male (\%77.7) & Female (\%22.3) & Conventional (\%71.6) & FSA/RHS (\%1.3) \\
\textbf{\textbf{(TV:0.03)}} & Non-White & Home Purchase (\%71.8) & Home Improv. (\%1.7) & Male (\%74.8) & Female (\%25.2) & Conventional (\%78.1) & FSA/RHS (\%0.3) \\ \midrule
\textbf{Sub-group 4} & White & Refinancing (\%61.8) & Home Improv. (\%9.2) & Male (\%65.8) & Female (\%34.2) & Conventional (\%78.5) & FSA/RHS (\%0.8) \\
\textbf{\textbf{(TV:0.1)}} & Non-White & Refinancing (\%65.9) & Home Improv. (\%7.7) & Male (\%62.9) & Female (\%37.1) & Conventional (\%79.9) & FSA/RHS (\%0.3)\\ \midrule

&  & \multicolumn{2}{c}{\textbf{Loan Purpose}} & \multicolumn{2}{c}{\textbf{Applicant Sex}} & \multicolumn{2}{c}{\textbf{Occupation Type}} \\ \cmidrule{3-8}
 &  & \textbf{Majority} & \textbf{Minority} & \textbf{Majority} & \textbf{Minority} & \textbf{Majority} & \textbf{Minority} \\ \midrule
 \multicolumn{8}{l} {\textbf{HDMA - Asian}}\\ \midrule
\textbf{Sub-group 1} & White & Refinancing (\%95.3) & Home Improv. (\%0.3) & Male (\%95.7) & Female (\%4.3) & Principal (\%98.1) & N/A (\%0.0) \\
\textbf{\textbf{(TV:0.03)}} & Non-White & Refinancing (\%92.3) & Home Improv. (\%0.1) & Male (\%96.2) & Female (\%3.8) & Principal (\%96.4) & Non-Principal (\%3.6) \\ \midrule
\textbf{Sub-group 2} & White & Refinancing (\%75.3) & Home Improv. (\%1.5) & Male (\%86.3) & Female (\%13.7) & Principal (\%90.8) & N/A (\%0.0) \\
\textbf{\textbf{(TV:-0.01)}} & Non-White & Refinancing (\%58.7) & Home Improv. (\%0.8) & Male (\%85.6) & Female (\%14.4) & Principal (\%86.2) & Non-Principal (\%13.8) \\ \midrule
\textbf{Sub-group 3} & White & Home Purchase (\%59.3) & Home Improv. (\%6.3) & Male (\%74.6) & Female (\%25.4) & Principal (\%90.4) & N/A (\%0.0) \\
\textbf{\textbf{(TV:0.02)}} & Non-White & Home Purchase (\%67.8) & Home Improv. (\%3.3) & Male (\%71.0) & Female (\%29.0) & Principal (\%82.9) & N/A (\%0.0) \\ \midrule
\textbf{Sub-group 4} & White & Refinancing (\%55.6) & Home Improv. (\%8.5) & Male (\%63.0) & Female (\%37.0) & Principal (\%94.1) & N/A (\%0.1) \\
\textbf{\textbf{(TV:0.08)}} & Non-White & Refinancing (\%62.5) & Home Improv. (\%5.7) & Male (\%58.0) & Female (\%42.0) & Principal (\%90.2) & Non-Principal (\%9.8)\\
 
\bottomrule
\hline
\end{tabular}}
\label{tab-adult}
\end{table*}

\subsection{Applications for fairness in Machine Learning}
In order to assess the practical utility and effectiveness of our causal disparity analysis in ML and automated decision-making, we trained an XGBoost classifier on both datasets to predict outcomes (referred to as the outcome node in Table~\ref{tab:dataset}). We created an 80-20\% train-test split using stratified sampling from all sub-groups (direct causal effect values). We computed classification results for the test set within each sub-group using AI Fairness 360~\cite{aif360-oct-2018} library. Tables~\ref{tab-ml} present the classification results, with the first row indicating the average performance and the last four rows representing the heterogeneity of performance across sub-groups. Across all experiments, performance varies among the sub-groups, with Sub-group 4 exhibiting worse performance and higher variability for all datasets except for the recall value for the Adult dataset; the Adult dataset (Precision: $0.76$(95\% CI interval: $0.016$), Recall: $0.71$($0.007$), and Accuracy:$0.74$($0.007$)) for HDMA-White (Precision:0.69(0.000), Recall: 0.89(0.013), Accuracy: 0.68(0.005)) and -Asian (Precision: 0.68(0.001), Recall:  0.86(0.011), Accuracy:  0.67(0.003). In total, the performance of the Sub-groups 1 and 4 are lower than the other Sub-groups.

\begin{figure*}[htb]
\centering
\begin{subfigure}{.33\textwidth}
    \centering
    \includegraphics[width=.99\linewidth]{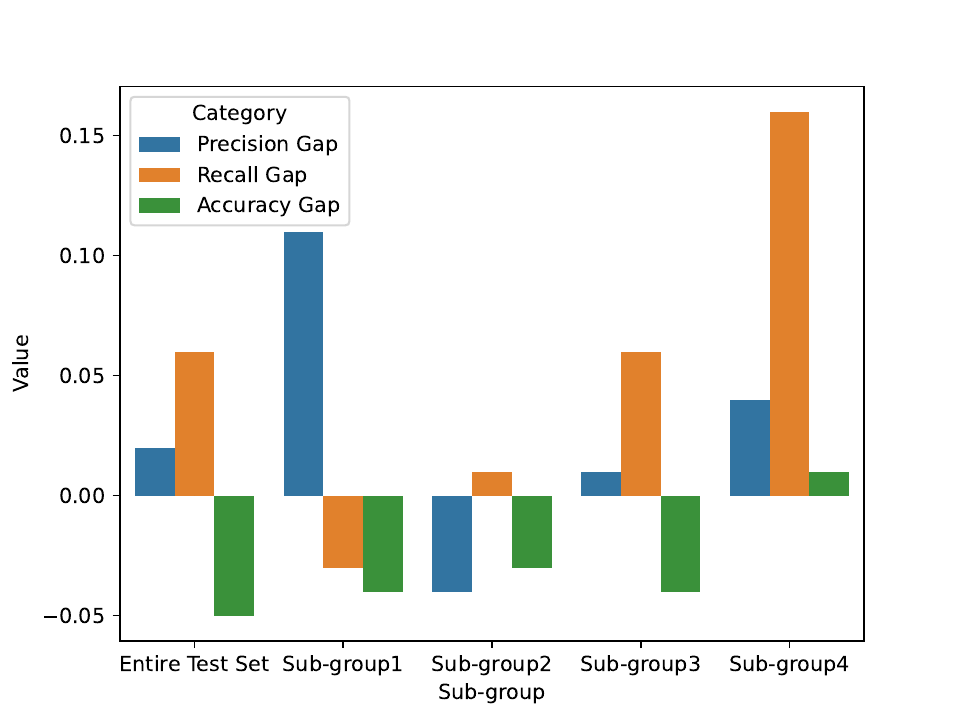}  
    \caption{Adult}
\end{subfigure}
\begin{subfigure}{.33\textwidth}
    \centering
    \includegraphics[width=.99\linewidth]{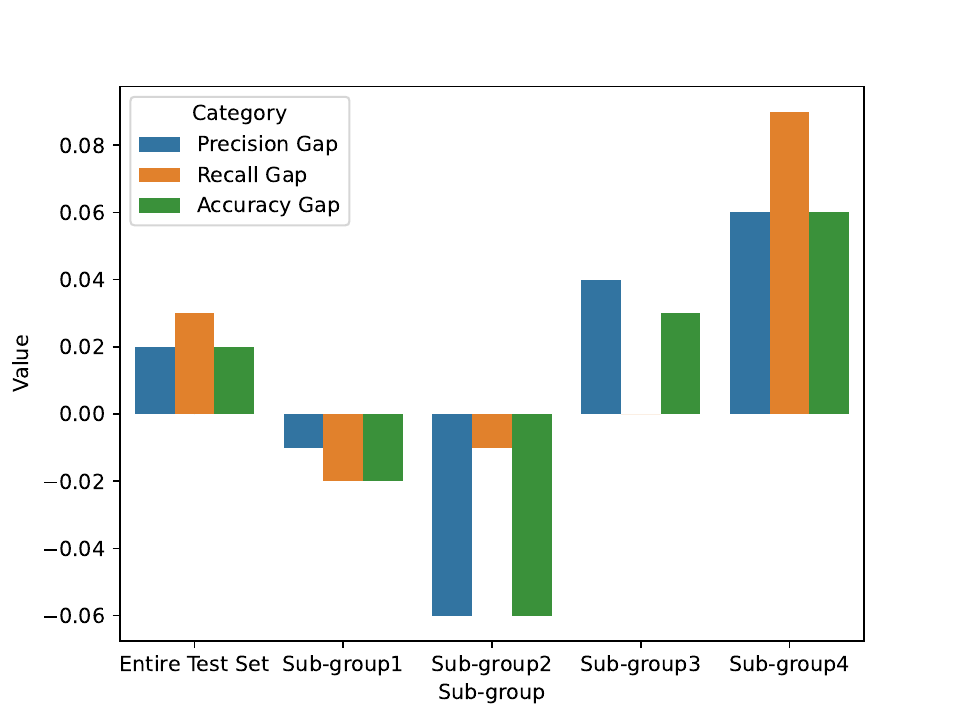}  
    \caption{HDMA-White}
    \label{fig:var_imp_hdma-w}
\end{subfigure}
\begin{subfigure}{.33\textwidth}
    \centering
    \includegraphics[width=.99\linewidth]{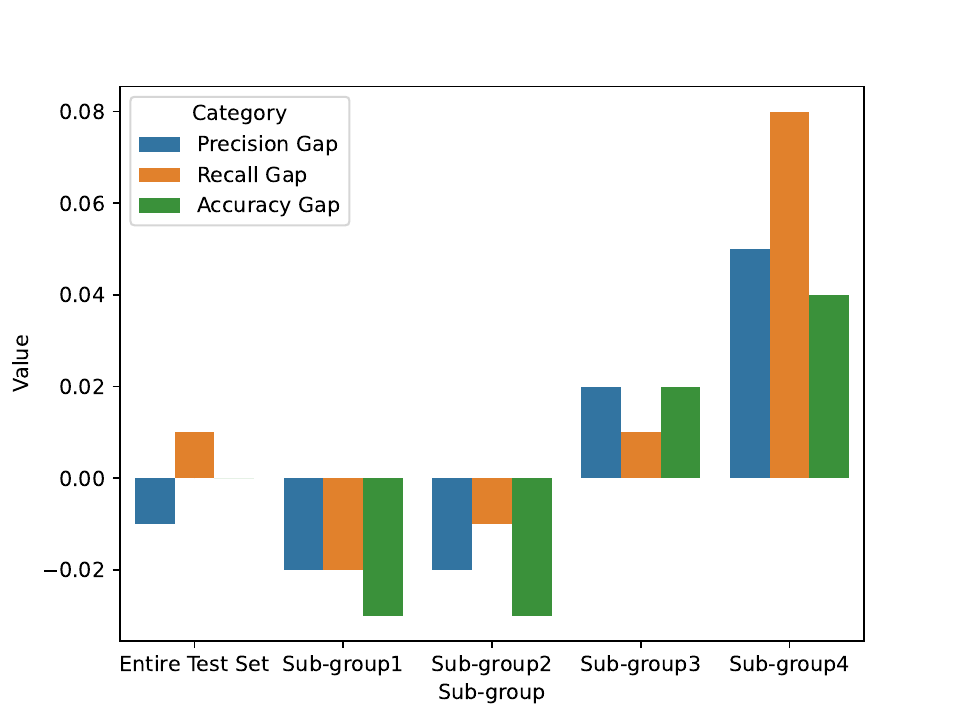}  
    \caption{HDMA-Asian}
    \label{fig:var_imp_hdma-a}
\end{subfigure}
\caption{Difference in performance metrics including accuracy, precision, and recall between sensitive groups.}
\label{fig:per-gap}
\end{figure*}

To better gauge the fairness of our ML classifier in our experiments and evaluate how decisions would differ if the circumstances were different, we plotted the performance gaps for the accuracy, recall, and precision between any two sensitive categories ($s_2$ - $s_1$) in Figure~\ref{fig:per-gap} across all sub-groups. Notably, almost 70\% of the performance gaps (positive gaps) favour the sensitive category $s_2$, which corresponds to the White individuals for Adult and HDMA-White, and the Non-Asian category for HDMA-Asian. As the plots indicate, the absolute value of the aggregated performance gap ranges from 0 to 0.07, whereas within sub-groups, the variation is more pronounced. The largest gaps are observed in Sub-group 1 (Precision:0.5, Recall:0.27, and Accuracy:0.29) for the Adult dataset and Sub-group 4 for HDMA-White (Precision:0.06, Recall:0.09, and Accuracy:0.06) and HDMA-Asian (Precision:0.05, Recall:0.08, and Accuracy:.04). The combination of performance gaps and lower performance in Sub-group 4 indicates the model's bias toward one of the sensitive categories. Of particular interest are the significant gaps in recall measures (higher false negative rates for one of the sensitive categories) among individuals in Sub-group 4 for both HDMA experiments. 

\begin{table*} [ht]\centering
\caption{The XGBoost model classification performance.} 
\resizebox{\textwidth}{!}{
\begin{tabular}{lrrrrrrrrrr}\toprule
\multirow{2}{*}{\textbf{}} &\multicolumn{3}{c}{\textbf{Adult} (mean (95\% CI interval))} &\multicolumn{3}{c}{\textbf{HDMA-White}(mean(95\% CI interval))} &\multicolumn{3}{c}{\textbf{HDMA-Asian}(mean(95\% CI interval))} \\\cmidrule{2-10}
&\textbf{Precision } &\textbf{Recall } &\textbf{Accuracy} &\textbf{Precision } &\textbf{Recall } &\textbf{Accuracy} &\textbf{Precision } &\textbf{Recall } &\textbf{Accuracy} \\\midrule

\textbf{Entire Test Set} &	0.77(0.012)&	0.66(0.004)&	0.87(0.003)&
0.78(0.001)&	0.95(0.006)&	0.76(0.003)&
0.78(0.001)&	0.96(0.005)&	0.77(0.002)\\
\textbf{Sub-group 1} & 0.76(0.016)&	0.61(0.008)&	0.77(0.007)&		
0.77(0.002)&	0.99(0.001)&	0.77(0.002)&	
0.78(0.001)&	0.99(0.005)&	0.77(0.003)\\
\textbf{Sub-group 2} & 0.78(0.021)&	0.63(0.004)&	0.93(0.002)&	    
0.82(0.001)&	0.99(0.003)&	0.81(0.001)&
0.82(0.001)&	0.99(0.004)&	0.81(0.002)\\
\textbf{Sub-group 3}     & 0.78(0.012)&	0.66(0.003)&	0.87(0.003)&	
0.83(0.001)&	0.99(0.004)&	0.82(0.001)&	
0.79(0.001)&	0.96(0.004)&	0.77(0.001)\\
\textbf{Sub-group 4}     &0.76(0.009)&	0.71(0.007)&	0.74(0.007)&	
0.69(0.000)&	0.89(0.013)&	0.68(0.005)& 
0.68(0.001)&	0.86(0.011)&	0.67(0.003)\\
\bottomrule
\end{tabular}}
\label{tab-ml}
\end{table*}

\section {Discussion}
\subsection{Main Findings}
In this study, we have demonstrated the utilization of causal disparity analysis to show the complex relationships and causal pathways linking sensitive attributes (such as \textit{race}) to real-world observational data outcomes (such as loan status or income) to supplement total variation (TV) also referred to as demographic parity. Our analysis is rooted in the assumptions of a basic causal graph, from which all findings are derived. Notably, our key finding reveals a direct causal link between \textit{race} and loan status or income, which might not have been apparent from the observed disparities alone. In the Adult dataset, our analysis reveals the presence of indirect effects through mediators, a phenomenon that resonates with prior research by Binkytė et al. \citep{binkyte2023causal}. However, the author's exploration of fairness measures across different causal discovery algorithms and causal paths demonstrated significant variability in the observed discrimination.

Considering the presence of direct causal effects within our datasets, we delved deeper into the variability among individuals regarding how \textit{race} directly influences their outcomes. This variability led to the identification of four distinct sub-groups, each sharing similar characteristics except for \textit{race}. In other words, within each sub-group, all covariates except \textit{race} remained consistent, with \textit{race} being hypothetically randomized. The ML model used in our study showed varying performance across these sub-groups. Sub-groups with higher and positive direct causal effects, which exhibited larger disparities in outcomes attributed to \textit{race}, experienced lower model performance. This performance gap within these sub-groups indicates potential unfairness and bias in the ML model, suggesting that \textit{race} may be a factor contributing to disparate outcomes. In all three experiments, the larger gap in false negative rates for Sub-group 4, which is not in favor of non-whites and Asians, suggests that the classifier tends to incorrectly predict loan status as rejected when it is actually accepted among these individuals, compared to white individuals within the same sub-group. This indicates a bias in the predictions against non-white individuals. Similarly, in the HDMA-Asian dataset, there is a similar disparity where predictions are biased against non-Asians. Furthermore, for the Adult dataset, in addition to the large recall gap for Sub-group 4, there is a large gap in the true positive rates in Sub-group 1 in favour of white individuals. This implies that the classifier is more successful at correctly predicting high income among white individuals in Sub-group 1 compared to non-white individuals within the same sub-group. This suggests a bias in favor of white individuals in predicting high income.

In essence, this is a nuanced finding that cannot be captured solely by dividing the entire sample size into privileged and unprivileged groups based on the sensitive attribute alone which is \textit{race} in our case. Our research findings are in accordance with existing literature in two significant respects. First, employing decomposed and structural causal analysis, our results resonate with a substantial body of research delving into mediating mechanisms by estimating both natural direct and indirect effects within the potential outcome framework \citep{jackson2021meaningful,park2022estimation,jeffries2019methodological}. Our causal methodology experiments echo the trajectory of research pioneered by counterfactual causal fairness analysis \citep{plecko2022causal} working on quantifying discrimination, decomposing variations, and deriving empirical measures of fairness from data. Second, considering heterogeneity in causal effects, our approach and findings align with other studies where the concept of heterogeneous treatment effects and the use of causal forest have been employed \citep{dandl2024makes}. For instance, similar methodologies have been leveraged in analyzing environmental policy effects \citep{miller2020causal}, conducting cost-effectiveness analyses encompassing outcomes, costs, and net monetary benefits \citep{bonander2021using}, as well as in assessing educational interventions and grading discrimination \citep{jin2019heterogeneous}.

\subsection{Limitations and Future Directions}
As ML advances at an unprecedented pace, its societal implications have attracted heightened scrutiny. Consequently, the importance of conducting disparity analysis has been emphasized in the contemporary landscape. While this study has provided valuable insights into causal disparity analysis, it's essential to acknowledge several limitations and explore potential avenues for improvement. The analysis primarily focused on disparities related to a single protective attribute, such as \textit{race}. However, this narrow focus may not fully capture the intricate interplay of multiple factors contributing to discrimination and bias in real-world scenarios. Future research should consider incorporating intersectional disparity analysis, which examines how multiple protective attributes intersect and interact to shape outcomes. In line with this, future work should also involve a thorough exploration of diverse causal discovery algorithms and identification methods. It's worth noting that the reliance solely on a basic causal graph framework in this study presents a limitation, as it may oversimplify the intricate causal relationships inherent in real-world data. Additionally, the datasets used in this study may not comprehensively represent the diversity and complexity of real-world populations. Limited diversity within the datasets can lead to biased results and may not encompass the full range of experiences and challenges faced by individuals from marginalized or underrepresented groups. Future work should involve utilizing more diverse and representative datasets, validating the findings within specific contexts, and identifying any context-specific factors that may influence fairness and bias.

To conclude, our study emphasized the imperative of delving into causal pathways, decomposing them, and assessing heterogeneity among individuals. This approach not only offers a comprehensive understanding of disparities within the data but also enables targeted interventions and strategies to promote fairness and equity.

\newpage
\clearpage
\bibliography{aaai25}
\appendix
\newpage

\section{Natural Effects}

In Table \ref{tab:natural-effects}, we provide estimations of natural effects using three methods—CFA-CRF \citep{plecko2022causal}, CFA-MedDML \citep{plecko2022causal}, and twangmediation \citep{coffman2023weighting}—for all three datasets.

\begin{table}[!b]\centering
\caption{Summary of natural effect measurments. Each value in the table is formatted as mean (standard deviation).}
\label{tab:natural-effects}
\resizebox{\textwidth}{!}{
\begin{tabular}{llllllll}\toprule
&\multicolumn{2}{c}{White} &\multicolumn{2}{c}{HDMA-White} &\multicolumn{2}{c}{HDMA-Asian} \\\cmidrule{2-7}
&NDE &NIE &NDE &NIE &NDE &NIE \\\midrule
CFA-CRF \citep{plecko2022causal} &0.016 (0.0001) &0.034 (0.0002) &0.062 (0.0001) &-0.01 (0.0001) &0.041 (0.0001) &-0.016 (0.0001) \\
CFA-MedDML \citep{plecko2022causal} &0.006 (0.0051) &0.044 (0.0019) &0.059 (0.0026) &-0.005 (0.0003) &0.041 (0.003) &-0.011 (0.0004) \\
twangmediation \citep{coffman2023weighting} &0.017 (0.008) &0.032 (0.001) &0.065 (0.003) &-0.007 (0.007) &0.048 (0.004) &-0.179 (0.492) \\
\bottomrule
\end{tabular}}
\end{table}

\section{Histogram of ctf-DE and NDE Values}

\begin{figure}[!hb]
\centering

\begin{subfigure}{.33\textwidth}
    \centering
    \includegraphics[width=.99\linewidth]{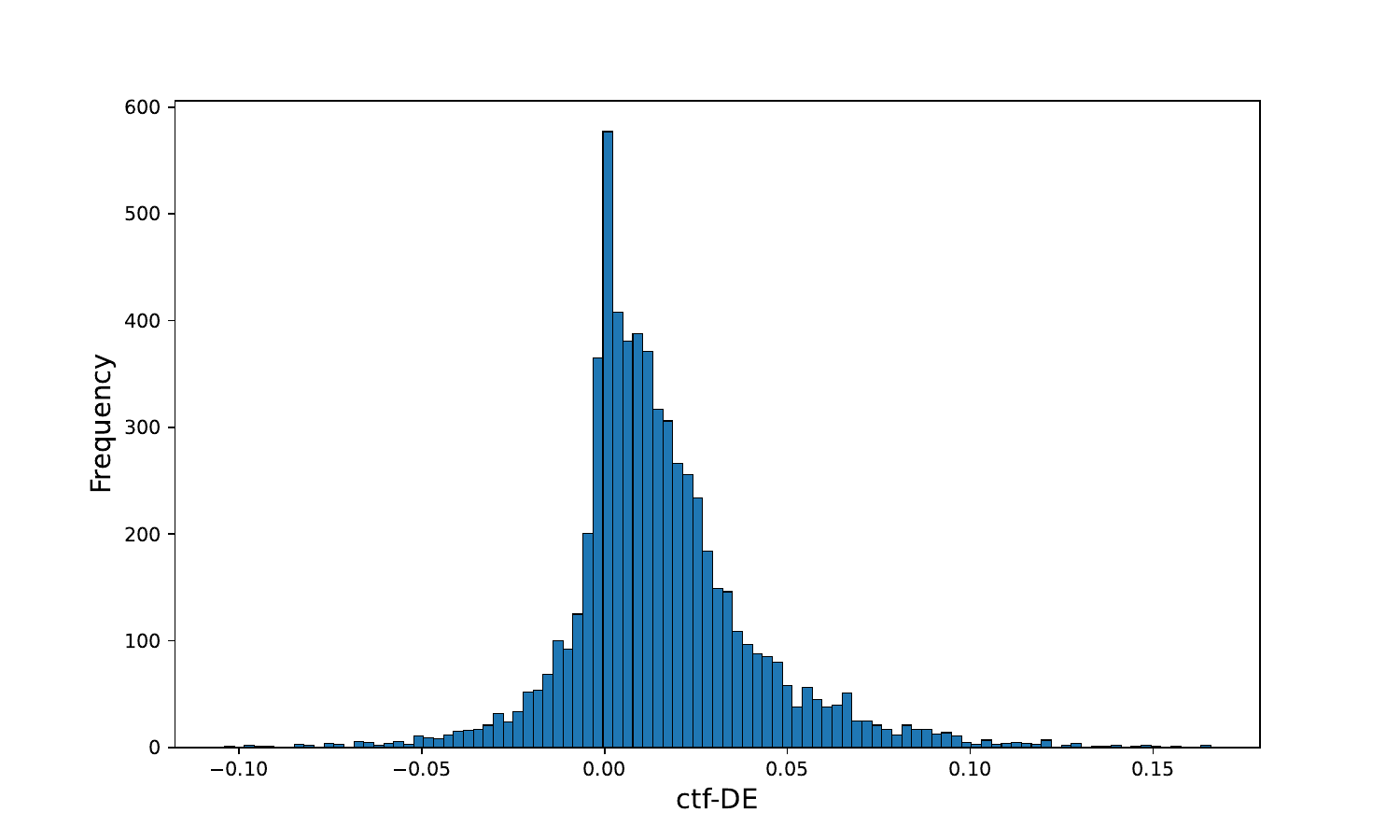}
    \caption{Adult dataset}
    \label{fig:adult-ctf-de-hist}
\end{subfigure}

\begin{subfigure}{.33\textwidth}
    \centering
    \includegraphics[width=.99\linewidth] {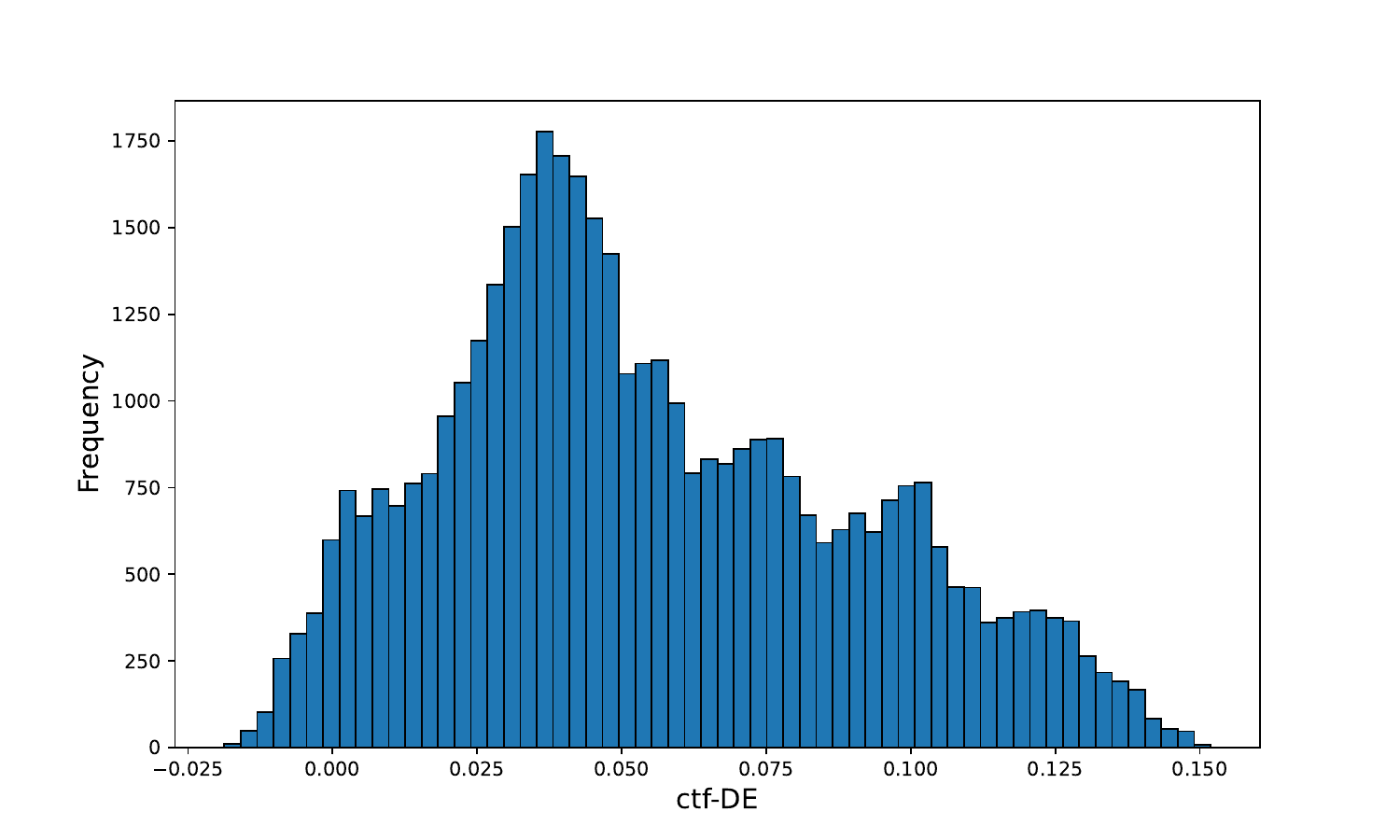}
    \caption{HDMA-White dataset}
\end{subfigure}
\begin{subfigure}{.33\textwidth}
    \centering
    \includegraphics[width=.99\linewidth]{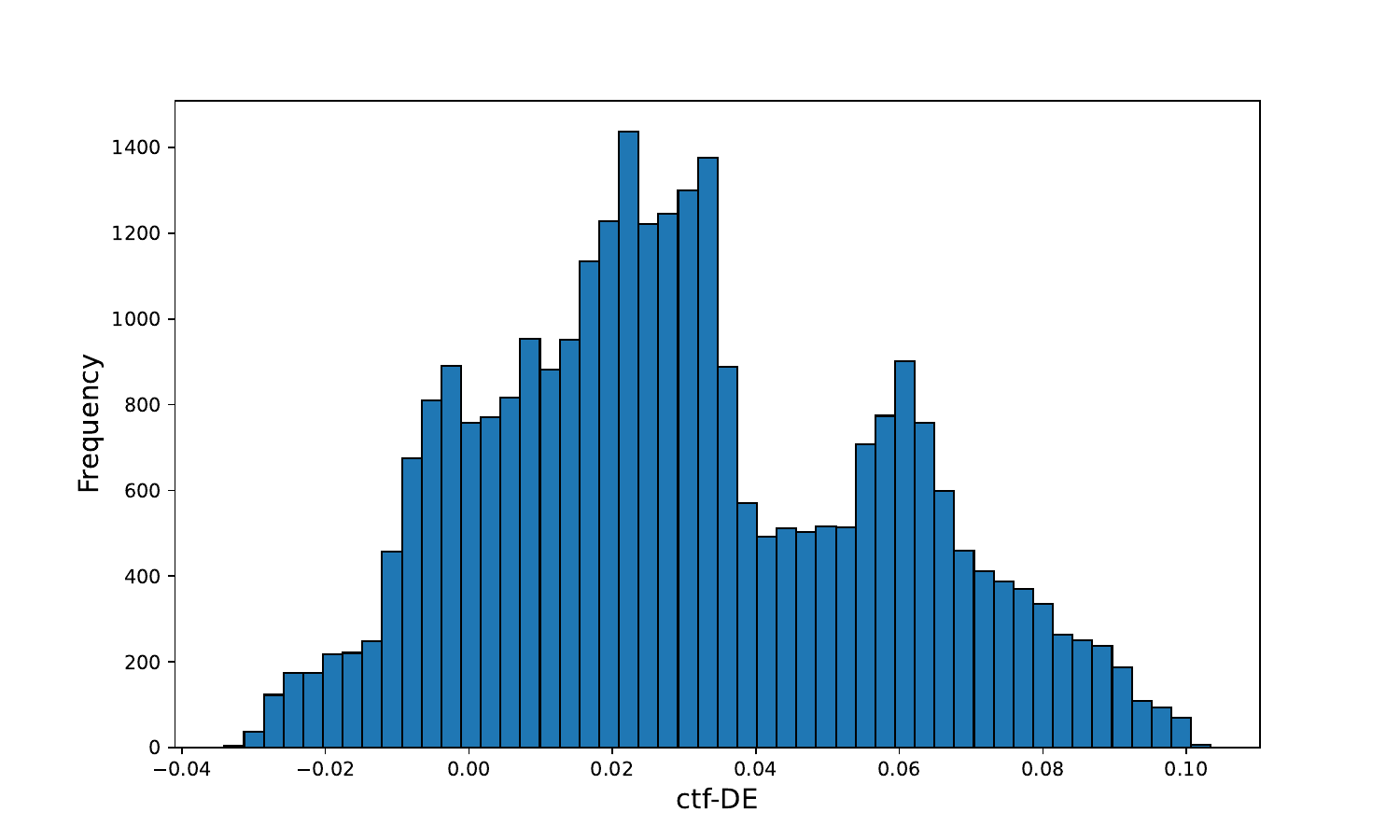}
    \caption{HDMA-Asian dataset}
    \label{fig:hdma-asian-ctf-de-hist}
\end{subfigure}
\caption{Histogram of ctf-DE values}
\label{fig:ctf_de_hist}
\end{figure}

\begin{figure}[!ht]
\centering
\begin{subfigure}{.33\textwidth}
    \centering
    \includegraphics[width=.99\linewidth]{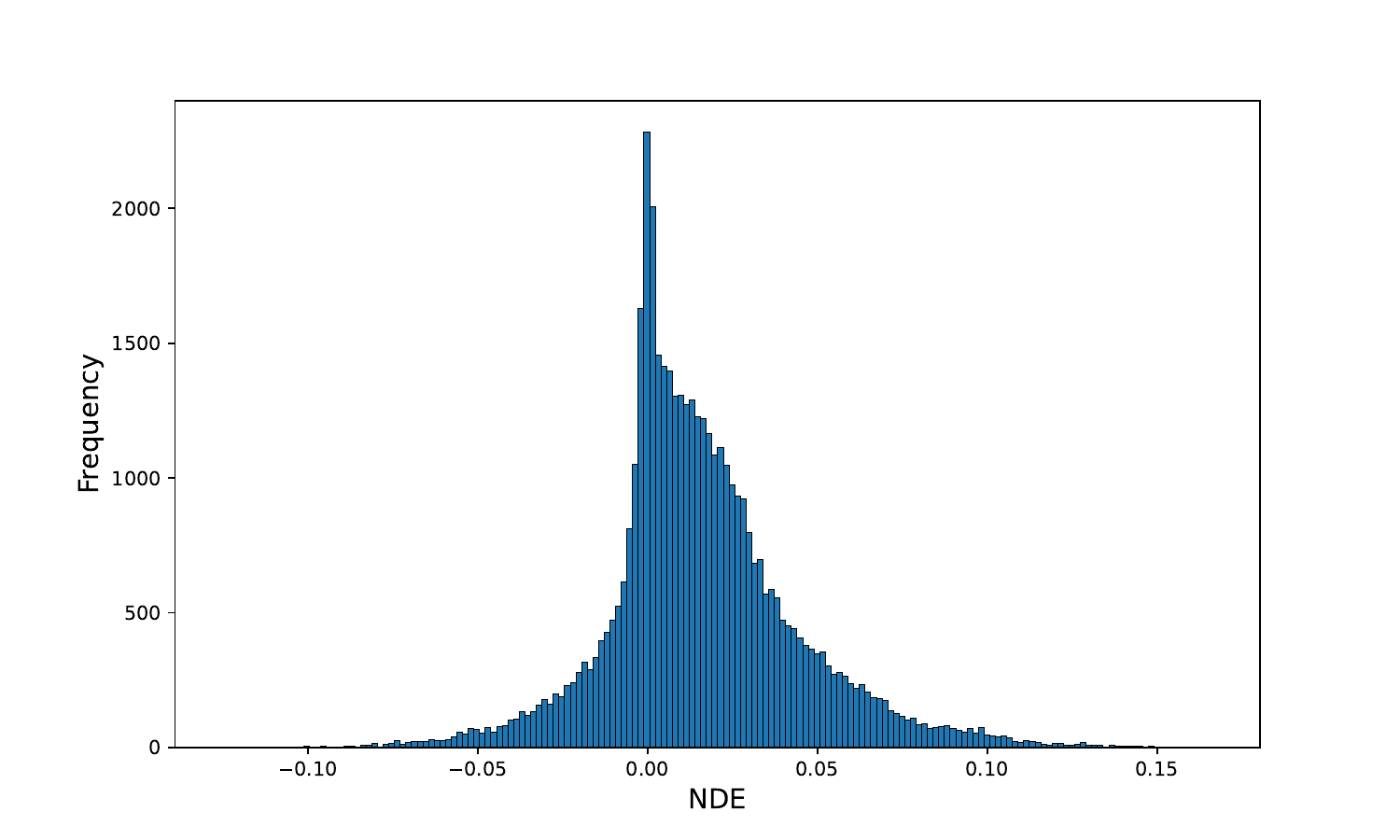}
    \caption{Adult dataset}
\end{subfigure}
\begin{subfigure}{.33\textwidth}
    \centering
    \includegraphics[width=.99\linewidth] {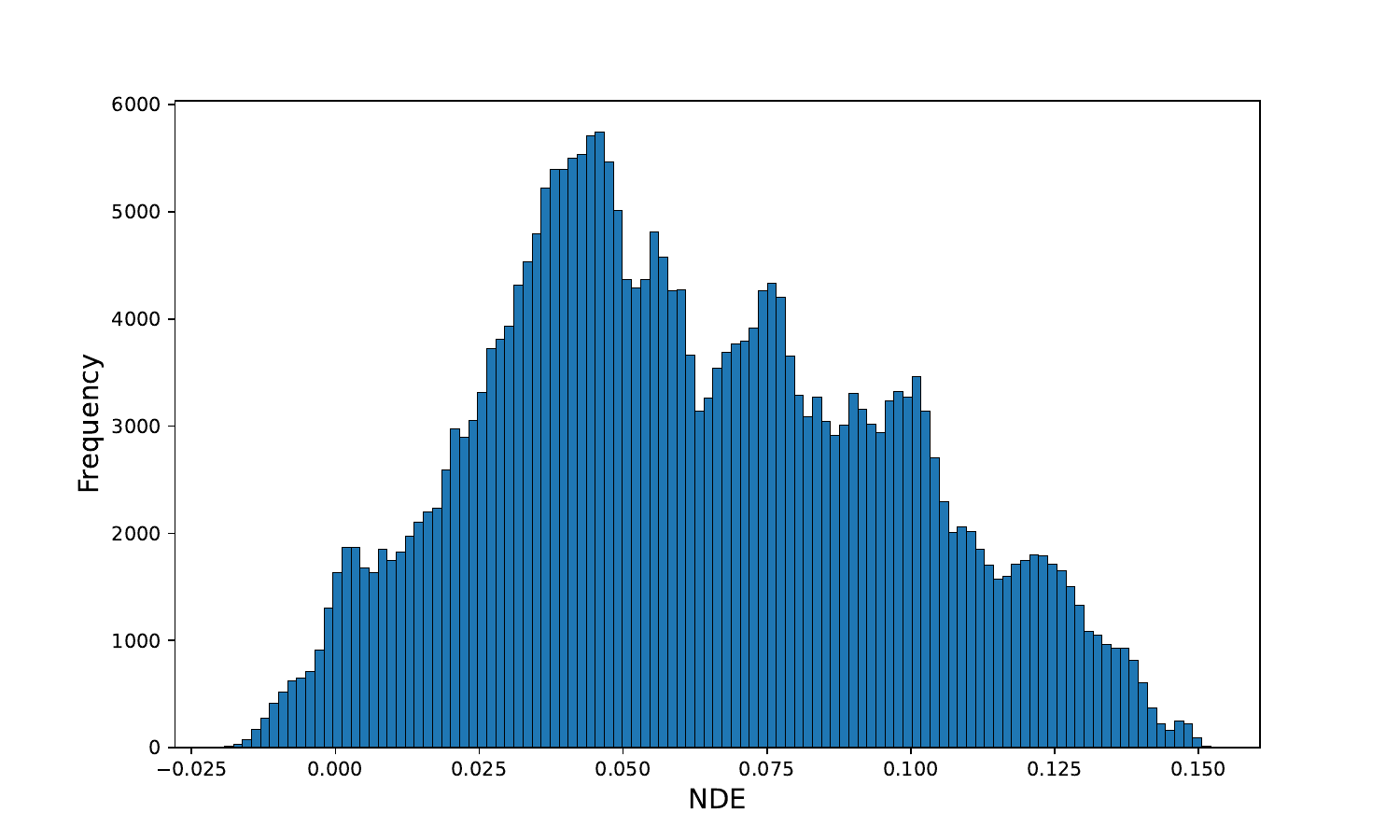}
    \caption{HDMA-White dataset}
\end{subfigure}
\begin{subfigure}{.33\textwidth}
    \centering
    \includegraphics[width=.99\linewidth]{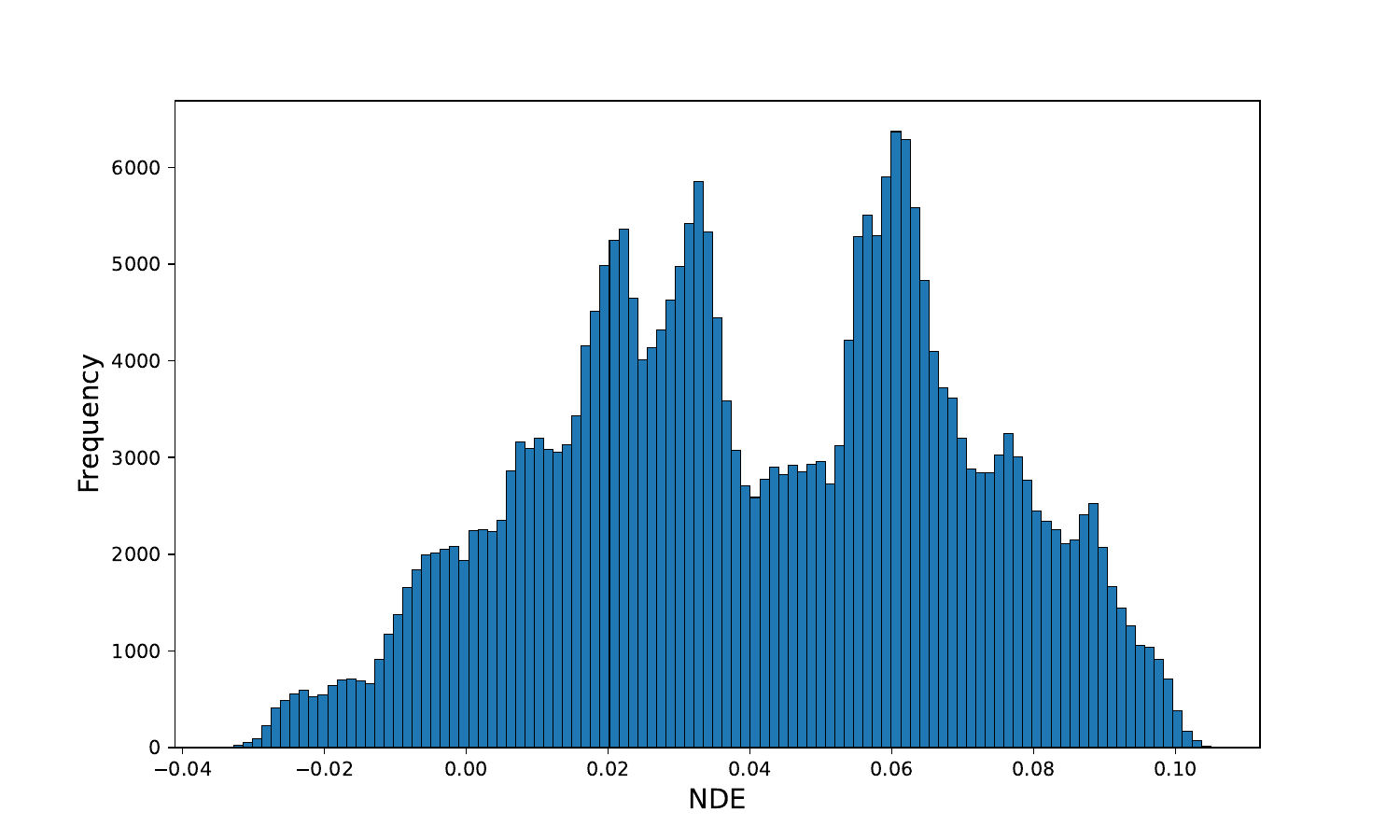}
    \caption{HDMA-Asian dataset}
\end{subfigure}
\caption{Histogram of NDE values}
\label{fig:nde_hist}
\end{figure}

In Figure \ref{fig:ctf_de_hist}, we provide histogram plots of ctf-DE values for the three datasets: Adult, HDMA-White, and HDMA-Asian. Each histogram provides a visual representation of the distribution and spread of ctf-DE values within each dataset. These figures provide us with the knowledge to find optimal sub-groups.
In Figure \ref{fig:nde_hist}, we provide histogram plots of NDE values for all three datasets as well.

\end{document}